\documentclass[10pt, conference]{IEEEtran}

\usepackage{url}
\usepackage{amsmath}
\usepackage{graphicx}
\usepackage{subfigure}
\usepackage{booktabs}
\usepackage{color}
\usepackage{multirow}
\usepackage{algorithm}
\usepackage{algpseudocode}
\usepackage{soul}
\usepackage{todonotes}
\usepackage[caption=false,font=normalsize,labelfont=sf,textfont=sf]{subfig}
\usepackage[belowskip=-1pt,aboveskip=0pt]{caption}
\usepackage{float}
\usepackage{epstopdf}
\usepackage{algpseudocode}
\usepackage{lipsum}

\usepackage{subfig}

\let\OLDthebibliography\thebibliography
\renewcommand\thebibliography[1]{
  \OLDthebibliography{#1}
  \setlength{\parskip}{0pt}
  \setlength{\itemsep}{0pt plus 0.3ex}
}

\begin{document}

\title{\huge{Storage and Memory Characterization of Data Intensive Workloads for Bare Metal Cloud  
}}
\author{Hosein Mohammadi Makrani$^1$ \\ 
\small {\em  $^1$George Mason University \quad
        } 
         \\ [2mm]
\small Submission Type: Research Draft
}
\maketitle

\begin{abstract}
As the cost-per-byte of storage systems dramatically decreases, SSDs are finding their ways in emerging cloud infrastructure. Similar trend is happening for main memory subsystem, as advanced DRAM technologies with higher capacity, frequency and number of channels are deploying for cloud-scale solutions specially for non-virtualized environment where cloud subscribers can exactly specify the configuration of underling hardware. Given the performance sensitivity of standard workloads to the memory hierarchy parameters, it is important to understand the role of memory and storage for data intensive workloads. In this paper, we investigate how the choice of DRAM (high-end vs low-end) impacts the performance of Hadoop, Spark, and MPI based Big Data workloads in the presence of different storage types on bare metal cloud. Through a methodical experimental setup, we have analyzed the impact of DRAM capacity, operating frequency, the number of channels, storage type, and scale-out factors on the performance of these popular frameworks. Based on micro-architectural analysis, we classified data-intensive workloads into three groups namely I/O bound, compute bound, and memory bound. The characterization results show that neither DRAM capacity, frequency, nor the number of channels play a significant role on the performance of all studied Hadoop workloads as they are mostly I/O bound. On the other hand, our results reveal that iterative tasks (e.g. machine learning) in Spark and MPI are benefiting from a high-end DRAM in particular high frequency and large number of channels, as they are memory or compute bound. Our results show that using SSD PCIe cannot shift the bottleneck from storage to memory, while it can change the workload behavior from I/O bound to compute bound.
\end{abstract}


\section{Introduction}
\label{sec:intro}

Advances in various branches of technology – data sensing, data communication, data computation, and data storage – are driving an era of unprecedented innovation for information retrieval. The world of big data is constantly changing and producing substantial amounts of data that creates challenges to process it using existing solutions. To address this challenge, several frameworks such as Hadoop and Spark, based on cluster computing, have been proposed. The main characteristic of these new frameworks is their ability to process large-scale data-intensive workloads on commodity hardware \cite{bertino2013big}.

Data analytics heavily rely on deep machine learning and data mining algorithms, and are running complex database software stack with significant interaction with I/O and OS, and exhibit high computational intensity and I/O intensity. In addition, unlike conventional CPU workloads, these workloads combine a high data rate requirement with high computational power requirement, in particular for real-time and near-time performance constraints.

Three well-known parallel programming frameworks used by community are Hadoop, Spark, and MPI. Hadoop and Spark are two prominent frameworks for big data analytics. Spark has been developed to overcome the limitation of Hadoop on efficiently utilizing main memory. Both Hadoop and Spark use clusters of commodity hardware to process large datasets. MPI, a de facto industry standard for parallel programming on distributed memory systems, is also used for data analytics \cite{wang2014bigdatabench}.

In the era of big data, it is important to evaluate the effect of memory and storage parameters on the performance of data intensive workloads. While there are literatures on understanding the behavior of those workloads, most of prior works have focused on the CPU parameters such as core counts, core frequency, cache parameters, and network configuration or I/O implication with the assumption of the demand for using the fastest and largest main memory in the commodity hardware \cite{dimitrov2013memory,alzuru2015hadoop,jia2013characterizing,pan2014characterization,liang2014performance,issa2016performance}.

Since the storage is a well-known bottleneck in data intensive workloads, several studies have been investigating the benefits of using Solid State Drives as a replacement for traditional Hard Disk Drive \cite{hong2016optimizing,islam2015triple,kim2011hybridstore,harter2014analysis,kambatla2014truth,krish2014venu,ahn2015analytical,choi2015early}. Solid-state storage offers several advantages over hard disks such as lower access latencies for random requests, higher bandwidth, and lower power consumption. Using SSD as storage raises the important question of whether the bottleneck has shifted to memory subsystem or not. To answer this question, we need to effectively characterize the performance of such workloads with respect to memory parameters. However, none of the previous works have studied the impact of main memory subsystem and storage at the same time to characterize data intensive workloads and the underlying frameworks.

The objective of this paper is to perform joint analysis of the impact of memory and storage parameters on the performance of data intensive workloads on bare metal cloud such as IBM/SoftLayer. To perform the memory subsystem analysis, we have investigated three configurable memory parameters including memory capacity, memory frequency, and number of memory channels, three types of storage (HDD, SSD SATA, and SSD PCIe), and scaling factors such as number of cores, core frequency, and number of nodes to determine how these parameters affect the performance of such workloads. This analysis helps in making architectural decision such as what memory architecture to use to build a server for data intensive workloads.

To the best of our knowledge this is the first work that looks beyond just the memory capacity to understand Hadoop, Spark and MPI based big data workloads’ memory and storage behavior by analyzing the effect of memory frequency as well as number of memory channels on the performance.

The remainder of this paper is organized as follows: Section 2 provides technical overview of the investigated workloads and the experimental setup. Results are presented in Section 3. Section 4 presents a discussion on the results. Section 5 describes related works. Finally, Section 6 concludes the paper.

\section{Experimental Setup}
\label{sec:method}

\subsection{Workloads}
For this study we target various domains of workloads namely that of microkernels, graph analytics, machine learning, E-commerce, social networks, search engines, and multimedia.  
We used BigDataBench \cite{wang2014bigdatabench} and HiBench \cite{huang2010hibench}. We selected a diverse set of workloads and frameworks to be representative of data intensive workloads. More details of these workloads are provided in Table \ref{tbl:tbl1}. The selected workloads have different characteristics such as high level data graph and different input/output ratios. Some of them have unstructured data type and some others are graph based. Also these workloads are popular in academia and are widely used in various studies.
In our study, we used Hadoop MapReduce version 2.7.1, Spark version 2.1.0 in conjunction with Scala 2.11, and MPICH2 version 3.2 installed on Linux Ubuntu 14.04. Our JVM version is 1.8. 

\begin{table*}[!htb] 
\caption{Studied workloads}
\centering
\scalebox{0.8}{
\begin{tabular}{|c|c|c|c|c|c|}
\hline
Workload & Domain & Input type & Input size (huge) & Framework & Suite \\
\hline
Wordcount & micro-kernel & text & 1.1 TB &  & \\
\cline{1-4}
Sort & micro-kernel & data & 178.8 GB & Hadoop, Spark, MPI & \\
\cline{1-4}
Grep & micro-kernel & text & 1.1 TB &  & BigData Bench \\
\cline{1-5}
Terasort & micro-kernel & data & 834 GB & Hadoop, Spark & \\
\cline{1-5}
Naive Bayes & E-commerce & Data & 306 GB & Hadoop, Spark, MPI & \\
\cline{1-5}
Page Rank & E-commerce & Data & 306 GB & Hadoop, Spark & \\
\hline
Bayes & E-commerce & Data & 306 GB & Hadoop, Spark & HiBench\\
\hline
k-means & Machine learning & Graph & 112.2 GB & Hadoop, Spark, MPI & BigDataBench\\
\hline
nweight & Graph analytics & Graph & 176 GB & Spark & \multirow{4}{*}{HiBench}\\
\cline{1-5}
Aggregation & Analytical query & Data & 1.08 TB & \multirow{3}{*}{Hadoop} & \\
\cline{1-4}
Join & Analytical query & Data & 1.08 TB &  & \\
\cline{1-4}
Scan & Analytical query & Data & 1.08 TB &  & \\
\hline
B.MPEG & Multimedia & DVD stream & 437 GB & \multirow{6}{*}{MPI} & \multirow{6}{*}{BigDataBench}\\
\cline{1-4}
DBN & Multimedia & Images & MNIST Dataset &  & \\
\cline{1-4}
Speech recognition & Multimedia & Audio & 252 GB &  & \\
\cline{1-4}
Image segmentation & Multimedia & Images & 162 GB &  & \\
\cline{1-4}
SIFT & Multimedia & Images & 162 GB &  & \\
\cline{1-4}
Face detection & Multimedia & Images & 162 GB &  & \\
\hline
\end{tabular}
}
\label{tbl:tbl1}
\end{table*}

\subsection{Hardware platform}
We carefully selected our experimental platform to investigate the micro architectural effect on the performance of data intensive workloads to understand whether our observations remain valid for future architectures with enhanced micro architecture parameters or not. This includes analyzing the results when increasing the core count, cache size and processor operating frequency. This is important, as the results will shed light on whether in future architectures larger number of cores, higher cache capacity and higher operating frequency change memory behavior of data intensive workloads or not. Using the data collected from our experimental test setup, we will drive architectural conclusion on how these micro architecture parameters are changing DRAM memory behavior and therefore impacting performance of workloads. While network overhead in general is influencing the performance of studied applications and therefore the characterization results, for data intensive applications, as shown in a recent work \cite{ousterhout2015making}, a modern high speed network introduces only a small 2\% performance benefit. We therefore used a high speed 1 Gbit/s network to avoid making it a performance bottleneck for this study. Our NICs have two ports and we used one of them per node for this study. For running the workloads and monitoring statistics, we used a six-node standalone cluster with detailed characteristics presented in Table \ref{tbl:tbl2}. We used single socket servers in this study, in order to hide the NUMA effect (to understand DRAM-only impact). To have a comprehensive experiment we used different SDRAM memory modules and all modules are provided from the same vendor.

\begin{table}[!htb] 
\caption{Hardware Platform}
\centering
\scalebox{0.83}{
\begin{tabular}{|c|l|l|}
\hline
Hardware  & \multirow{2}{*}{Parameter} & \multirow{2}{*}{Value} \\
type & & \\
\hline
\multirow{13}{*}{CPU} & \multirow{2}{*}{Model} & Intel Xeon  \\
 & & E5-2683 V4 \\ 
 \cline{2-3}
 & \# Core & 	16 (32 thread) \\
 \cline{2-3}
& Base Frequency &	2.1 GHz \\
\cline{2-3}
& Turbo Frequency	& 3.0 GHz \\
\cline{2-3}
& TDP	& 120 \\
\cline{2-3}
& L3 Cache &	40 MB\\ 
\cline{2-3}
& Memory Type 	& DDR4 \\
 & Support & 1866/2133/2400 \\
 \cline{2-3}
& Maximum Memory  &	\multirow{2}{*}{76.8 GB/S} \\
& Bandwidth &	 \\
\cline{2-3}
& Max Memory  	& \multirow{2}{*}{4} \\
& Channels supported	&  \\
\hline
\multirow{3}{*}{Disk} & Model & Samsung 960 PRO M.2 \\ 
\cline{2-3}
& Capacity	& 512 GB \\
\cline{2-3}
(SSD PCIE) & Speed	& Max 3.5 GB/S \\
\hline
\multirow{3}{*}{Disk} & Model & HyperX FURY \\ 
\cline{2-3}
& Capacity	& 480 GB \\
\cline{2-3}
(SSD SATA) & Speed	& 500 MB/S \\
\hline
\multirow{3}{*}{Disk} & Model & Seagate  \\ 
\cline{2-3}
& Capacity	& 500 GB \\
\cline{2-3}
(HDD) & Speed	& 7200 RPM \\
\hline
Network & Model & ST1000SPEXD4 \\
\cline{2-3}
Interface card & Speed & 1000 Mbps \\ 
\hline
\end{tabular}
}
\label{tbl:tbl2}
\end{table}

\subsection{Methodology}
The experimental methodology of this paper is focused on understanding how studied frameworks are utilizing main memory and storage.

1) Data collection: We used Intel Performance Counter Monitor tool (PCM) \cite{pcm} to understand hardware (memory and processor) behavior. Several works used performance counters to estimate the performance and power consumption of processors \cite{aspdac,igsc-sayadi,iccd} or employed performance counter for enhancing the security \cite{cf2018,DAC2018,Trustcom2018}. In this work we use performance counters to study  the memory behavior. The performance counter data are collected for the entire run of each workload. We collect OS-level performance information with DSTAT tool—a profiling tool for Linux based systems by specifying the event under study. Some of the metrics that we used for this study are memory footprint, L2, and Last Level Cache (LLC) hits ratio, instruction per cycle (IPC), core C0 state residency, and power consumption. 

2) Parameter tuning: For both Hadoop and Spark frameworks, it is important to set the number of mapper and reducer slots appropriately to maximize the performance. Based on the result of \cite{ferdman2012clearing}, the maximum number of mappers running concurrently on the system to maximize performance should be equal to the total number of available CPU cores in the system. Therefore, for each experiment, we set the number of mappers equal to the total number of cores. We also follow same approach for the number of parallel tasks in MPI. Adjusting default memory parameters of Hadoop and Spark also is important. Hence, we tuned Hadoop and Spark memory related configuration parameters. Followings are two most important memory related parameters that we tuned for all experiments: 

\textit{mapreduce.map.memory.mb}: is the upper memory limit that Hadoop allows to be allocated to a mapper, in megabytes. 
\textit{spark.executor.memory}: Amount of memory to use per executor process in Spark (e.g. 2g, 8g).

We set those values according to the following (we reserved 20\% of DRAM capacity for OS):
\begin{equation}
\begin{aligned}
\centering
& mapreduce.map.memory.mb = \\ & (DRAM capacity \times 0.8) / \\ 
& Number of Concurrent Mappers per Node
\end{aligned}
\end{equation}

\begin{equation}
\begin{aligned}
\centering
&spark.executor.memory = \\ & ((DRAM capacity - spark.driver.memory) \times 0.8) / \\ 
& Number of Executor per Node
\end{aligned}
\end{equation}

A recent work has shown that among all tuning parameters in a MapReduce framework, HDFS block size is most influential on the performance \cite{malik2016characterizing}. HDFS Block size has a direct relation to the number of parallel tasks (in Spark and Hadoop), as shown in EQ. (\ref{eq:nt}). 

\begin{eqnarray}\label{eq:nt}
\begin{aligned}
\centering
Number of Tasks = Input Size / Block Size  
\end{aligned}
\end{eqnarray}

In the above equation, the input size is the size of data that is distributed among nodes. The block size is the amount of data that is transferred among nodes. Hence, block size has impact on the network traffic and its usage. Therefore, we first evaluate how changing this parameter affects the performance of the system. We studied a broad range of HDFS block sizes varying from 32 MB to 1GB when the main memory capacity is 64 GB and it has the highest frequency and number of channels. Table 5 demonstrates the best HDFS configuration for maximizing the performance in both Hadoop and Spark frameworks based on the ratio of Input data size to the total number of available processing cores, and the workload class. The rest of the experiments presented in this paper are based on Table \ref{tbl:tbl3} configuration. Our tuning methodology guarantees to put the highest pressure on memory subsystem. 

\begin{table}[!htb] 
\caption{HDFS block size tuning}
\centering
\scalebox{0.83}{
\begin{tabular}{|c|c|c|c|c|}
\hline
Application & \multicolumn{4}{|c|}{$Input\ size/(\# \ nodes \times \# cores\ per\  node$)} \\
\cline{2-5}
class & \textbf{$<$64 MB} & \textbf{$<$512 MB} & \textbf{$<$4 GB} & \textbf{$>$ 4 GB} \\
\hline 
CPU & \multirow{2}{*}{32 MB} & \multirow{2}{*}{64 MB} & \multirow{2}{*}{128 MB} &  \multirow{2}{*}{256 MB} \\
intensive & & & & \\
\hline
I/O & \multirow{2}{*}{64 MB} & \multirow{2}{*}{256 MB} & \multirow{2}{*}{512 MB} &  \multirow{2}{*}{1 GB} \\
intensive & & & & \\
\hline
Iterative & \multirow{2}{*}{64 MB} & \multirow{2}{*}{128 MB} & \multirow{2}{*}{256 MB} &  \multirow{2}{*}{512 MB} \\
tasks & & & & \\
\hline
\end{tabular}
}
\label{tbl:tbl3}
\end{table}

\begin{figure*}[!t]
\centering
 \includegraphics[width=1\textwidth]{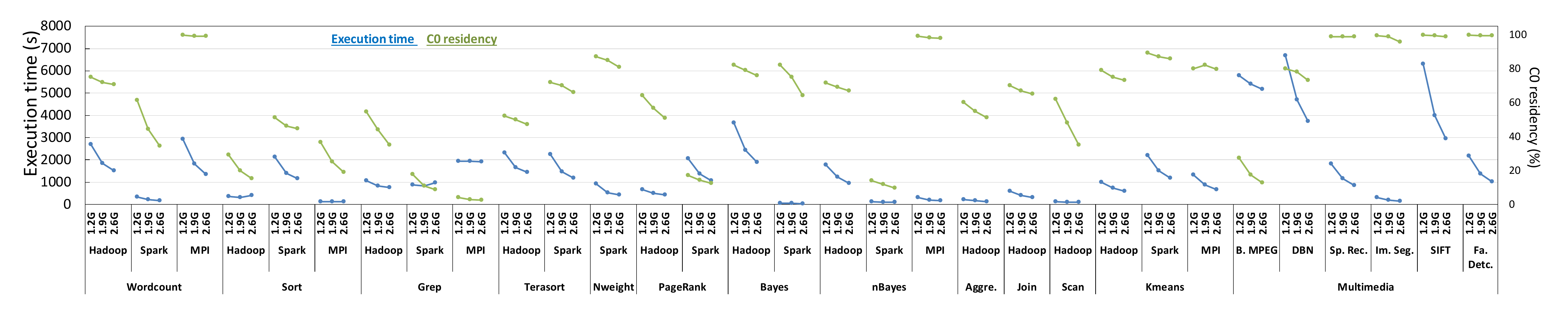}\\
\caption{Impact of CPU frequency on execution time and C0 state residency}
\label{fig:execution-c0}
\end{figure*}

\section{Results}
\label{sec:results}
 
Our experimental results are presented in this section. First, we present the classification of workloads into memory bound, compute bound, and I/O bound, which helps to accurately present the relation of performance and workload characteristics. Then, we present the memory analysis of the studied workloads. Then, we provide results of architectural implication of processor parameters on data intensive workloads. All performance metrics such as execution time, CPU active state residency, LLC and L2 hit ratio are discussed in this section. We also discuss the impact of storage system, and cluster size on memory subsystem. 

\subsection{Classification of workloads}

As the main goal of this paper is to study the combined impact of node architecture and data intensive workload's characteristics, it is important to classify those workloads first. To this goal, we have explored the micro architectural behavior of studied workloads to classify them and find more insights. Throughout this section we will present the results based on high speed SSD disk.

1) Core frequency implication: Figure \ref{fig:execution-c0} shows that studied workloads behave in two distinct ways. The execution time of the first group is decreased linearly by increasing the core frequency. The second group’s execution time does not drop significantly by increasing the CPU frequency, particularly when changing frequency from 1.9 GHz to 2.6 GHz. These two trends indicate that studied workloads have distinct behaviors of being either CPU bound or I/O bound. This conclusion further advocated by C0 state residency of processor. This proves sort, grep, PageRank, and scan from Hadoop, wordcount, grep, PageRank, Bayes, and nBayes from Spark, and sort, BasicMPEG, and grep from MPI to be Disk-intensive while others to be CPU-intensive. This can be explained as follows: If increasing the processor’s frequency reduces the active state residency (C0) of the processor, the workload is I/O bound, as when a core is waiting for I/O, the core changes its state to save power. Similarly, if active state residency does not change the workload is CPU bound.

\begin{figure}[t]
\centering
 \includegraphics[width=0.48\textwidth]{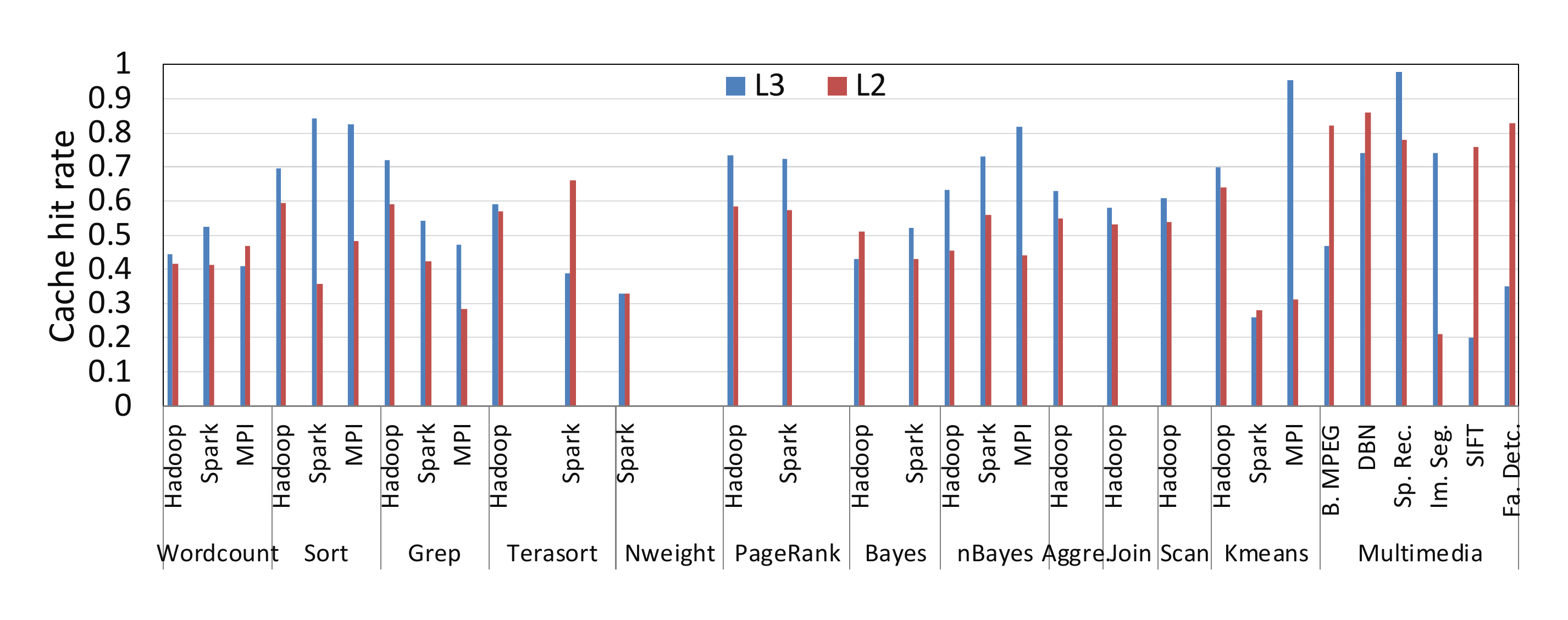}\\
\caption{LLC and L2 hit rate}
\label{fig:cache}
\end{figure}

2) Cache implication: Modern processor has a 3-level cache hierarchy.  Figure \ref{fig:cache} shows cache hit ratio of level 2 (L2) and last level cache (LLC). The results reveal an important characteristic of data intensive workloads. Our experimental results show most of the studied workloads (particularly MapReduce workloads) have a much higher cache hit ratio, which helps reducing the number of accesses to the main memory. Based on simulation as well as real-system experiment results in recent works, it is reported  that these applications' cache hit rate is too low (under 10\%) \cite{dimitrov2013memory,alzuru2015hadoop} for a system with 10 MB of LLC and for having LLC hit rate of 40\%, the system should have around 100 MB of LLC. However, our real system experimental results show that most of data intensive workloads have much higher LLC hit rate (more than 50\%) with only 40 MB LLC. The reason of high cache hit ratio is that each parallel task of MapReduce framework processes data in a sequential manner. This behavior increases the cache hits; therefore it prevents excessive access to DRAM.  Hence, based on the cache hit ratio of CPU intensive workloads and the intensity of accesses to memory, we can classify CPU bound workloads into two more groups namely compute bound and memory bound. If the cache hit ratio is low and the workload is an iterative task, it is classified as memory intensive. Our characterization showed that N-weight and Kmeans from Spark, and Image Segmentation from MPI are memory intensive. 
Therefore, we divided our workloads into three major groups of I/O bound, compute bound, and memory bound. Based on this classification we present our result in the following sections.

\subsection{Memory analysis}
In this section, we present a comprehensive discussion on memory analysis results to help better understanding the memory requirements of data intensive Workloads.

1) Memory channels implication: The off-chip peak memory bandwidth equation is shown in EQ. (\ref{eq:bw}).
\begin{eqnarray}\label{eq:bw}
\begin{aligned}
\centering
Bandwidth = Channels \times Frequency \times Width  
\end{aligned}
\end{eqnarray}

\begin{figure}[t]
\centering
 \includegraphics[width=0.48\textwidth]{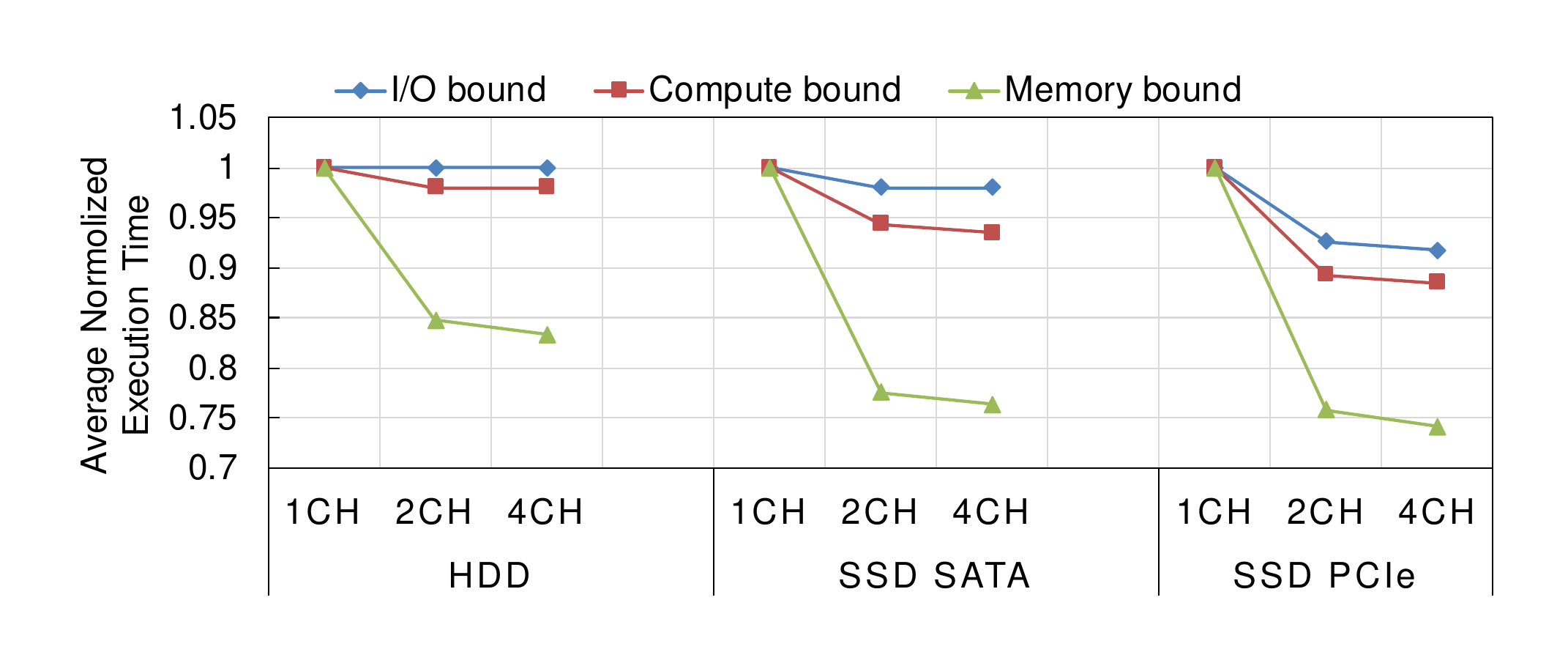}\\
\caption{Effect of channel on the execution time (Normalized to 1CH)}
\label{fig:ch}
\end{figure}

\begin{figure}[t]
\centering
 \includegraphics[width=0.48\textwidth]{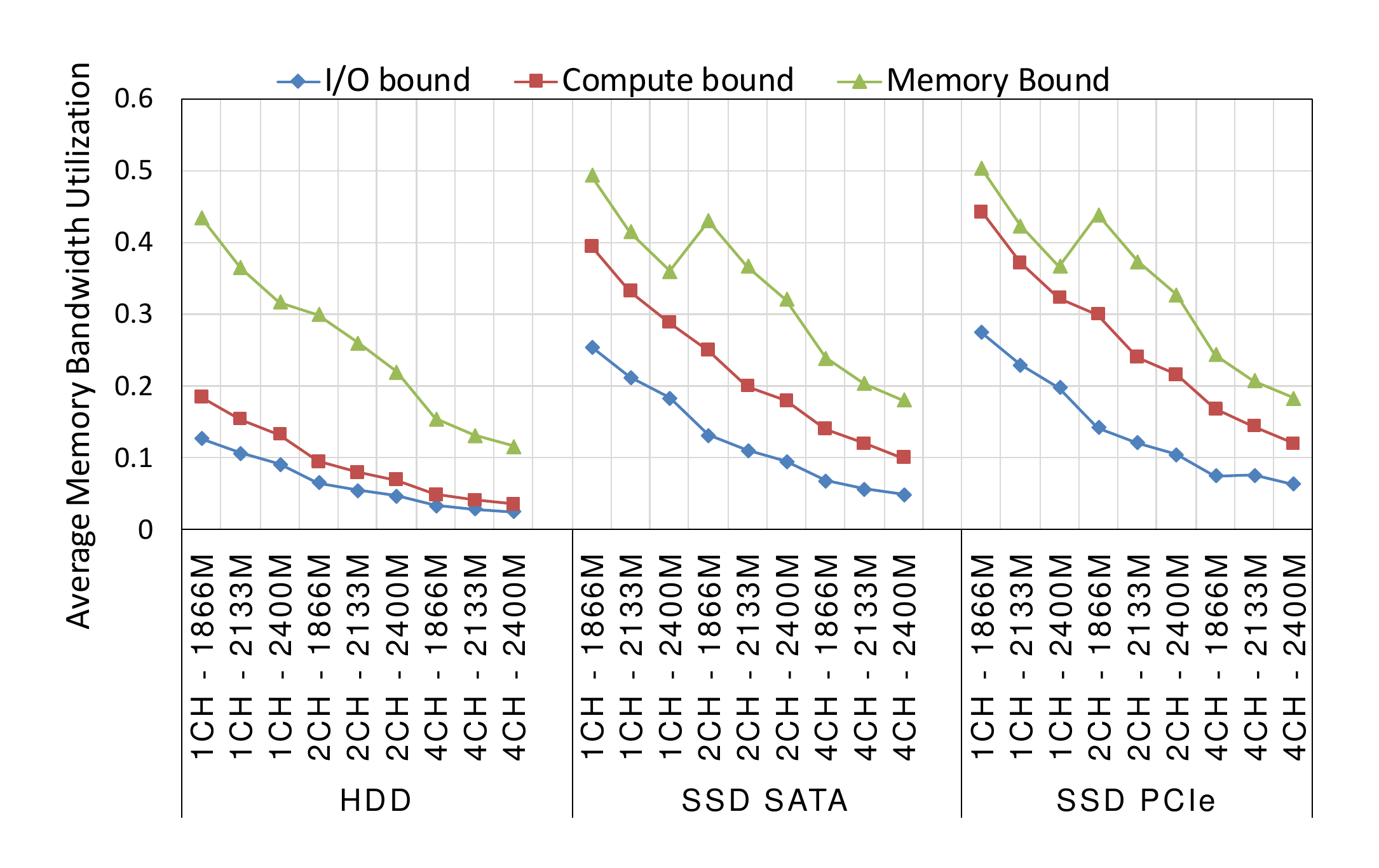}\\
\caption{Average memory bandwidth utilization}
\label{fig:util}
\end{figure}

We observe in Figure \ref{fig:ch} that increasing the number of channels have significant effect on the execution time of memory bound workloads (All of them are iterative tasks). Figure \ref{fig:util} provides more insights to explain this exceptional behavior. This Figure demonstrates the memory bandwidth utilization of each group. Bandwidth utilization of memory bound workloads is shown to be substantially higher than other workloads. Hence providing more bandwidth decreases their execution time. By increasing the number of channels from 1 to 4, the gain is found to be 28\%.   

\begin{figure}[t]
\centering
 \includegraphics[width=0.48\textwidth]{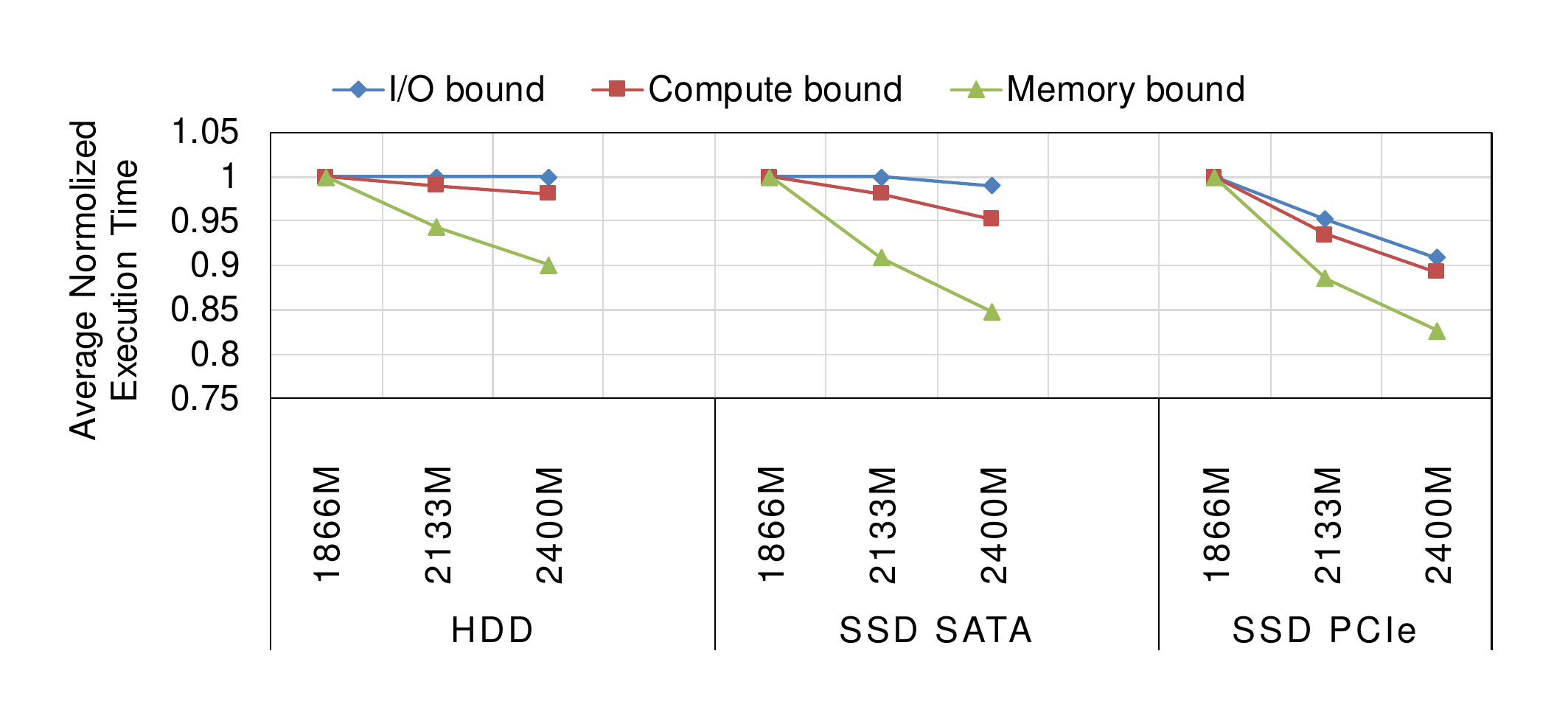}\\
\caption{Effect of memory frequency on the execution time (Normalized to 1866M)}
\label{fig:memfr}
\end{figure}

2) Memory frequency implication: As results in Figure \ref{fig:memfr} shows, similarly we do not observe significant improvement of bandwidth utilization or execution time by increasing memory frequency (from 1866 MHz to 2400 MHz) for none memory bound workloads. This finding may mislead to use the lowest memory frequency for other workloads. Based on EQ. (\ref{eq:rl}), read latency of DRAM depends on the memory frequency. 
\begin{eqnarray}\label{eq:rl}
\begin{aligned}
\centering
Read latency = 2 \times (CL / Frequency) 
\end{aligned}
\end{eqnarray}

However, for DDRx (e.g. DDR3), this latency is set fixed by the manufacturer with controlling CAS latency (CL). This means two memory modules with different frequency (1333 MHz and 1866 MHz) and different CAS Latency (9 and 13) can have the same read latency of 13.5 ns, but provide different bandwidth per channel (10.66 GB/s and 14.93 GB/s). Hence, as along as reduction of frequency does not change the read latency, it is recommended to reduce DRAM frequency for most of data intensive applications unless the application is memory intensive. Later in this paper we will discuss memory sensitivity of studied applications.

\begin{figure}[t]
\centering
 \includegraphics[width=0.48\textwidth]{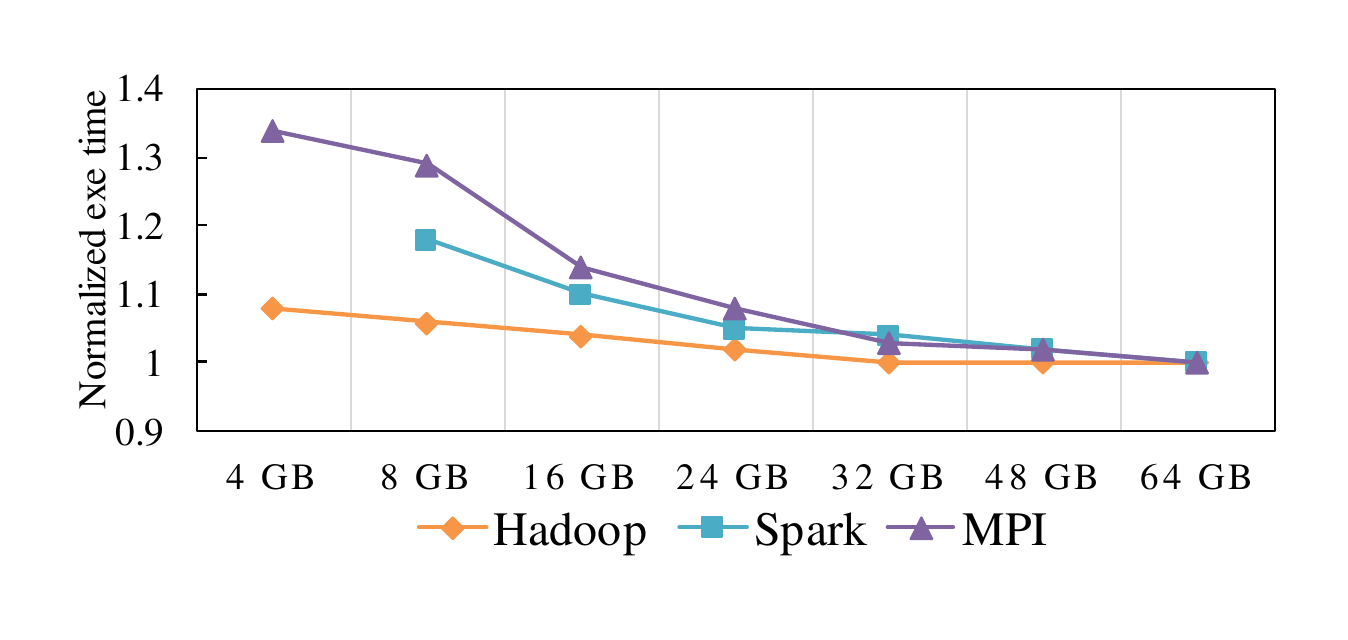}\\
\caption{Impact of memory capacity per node on performance}
\label{fig:capacity}
\end{figure}

\begin{figure}[t]
\centering
 \includegraphics[width=0.48\textwidth]{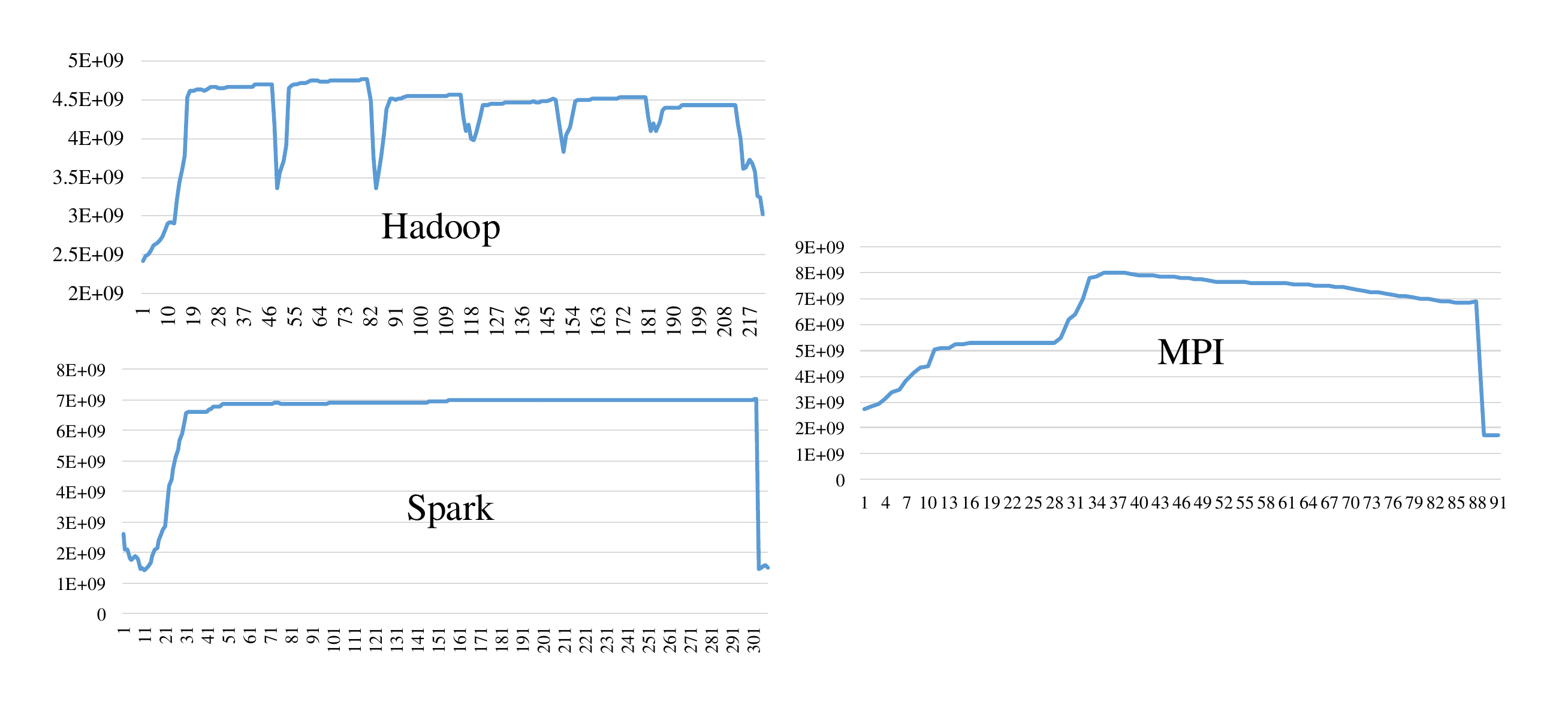}\\
\caption{K-means memory usage on various frameworks}
\label{fig:musage}
\end{figure}

3) DRAM capacity implication: To investigate the effect of memory capacity on the performance of data intensive workloads, we run all workloads with 7 different memory capacities per node. During our experiments, Spark workloads encountered an error when running on a 4GB memory capacity per node due to lack of memory space for the Java heap. Hence, the experiment of Spark workloads is performed with at least 8 GB of memory. Based on our observation, we found that only MPI workloads have an unpredictable memory capacity usage. In fact, a large memory capacity has no significant effect on the performance of studied Hadoop and Spark workloads. Hadoop workloads do not require high capacity memory because Hadoop stores all intermediate values generated by map tasks on the storage. Hence, regardless of the number of map tasks or input size, the  memory usage remains almost the same. Spark uses RDD to cache intermediate values in memory. Hence, by increasing the number of map tasks to run on a node, the memory usage increases. Therefore, by knowing the number of map tasks assigned to a nodes and the amount of intermediate values generated by each task, the maximum memory usage of Spark workloads is predictable per node. To better understand the impact of memory capacity on the performance, we have provided the average normalized execution time of these three frameworks in Figure \ref{fig:capacity} (Normalized to 64 GB). To illustrate how these frameworks utilize DRAM capacity we present K-means memory usage on 3 different frameworks in Figure \ref{fig:musage}. 

\subsection{Performance  analysis}
\begin{figure}[t]
\centering
 \includegraphics[width=0.48\textwidth]{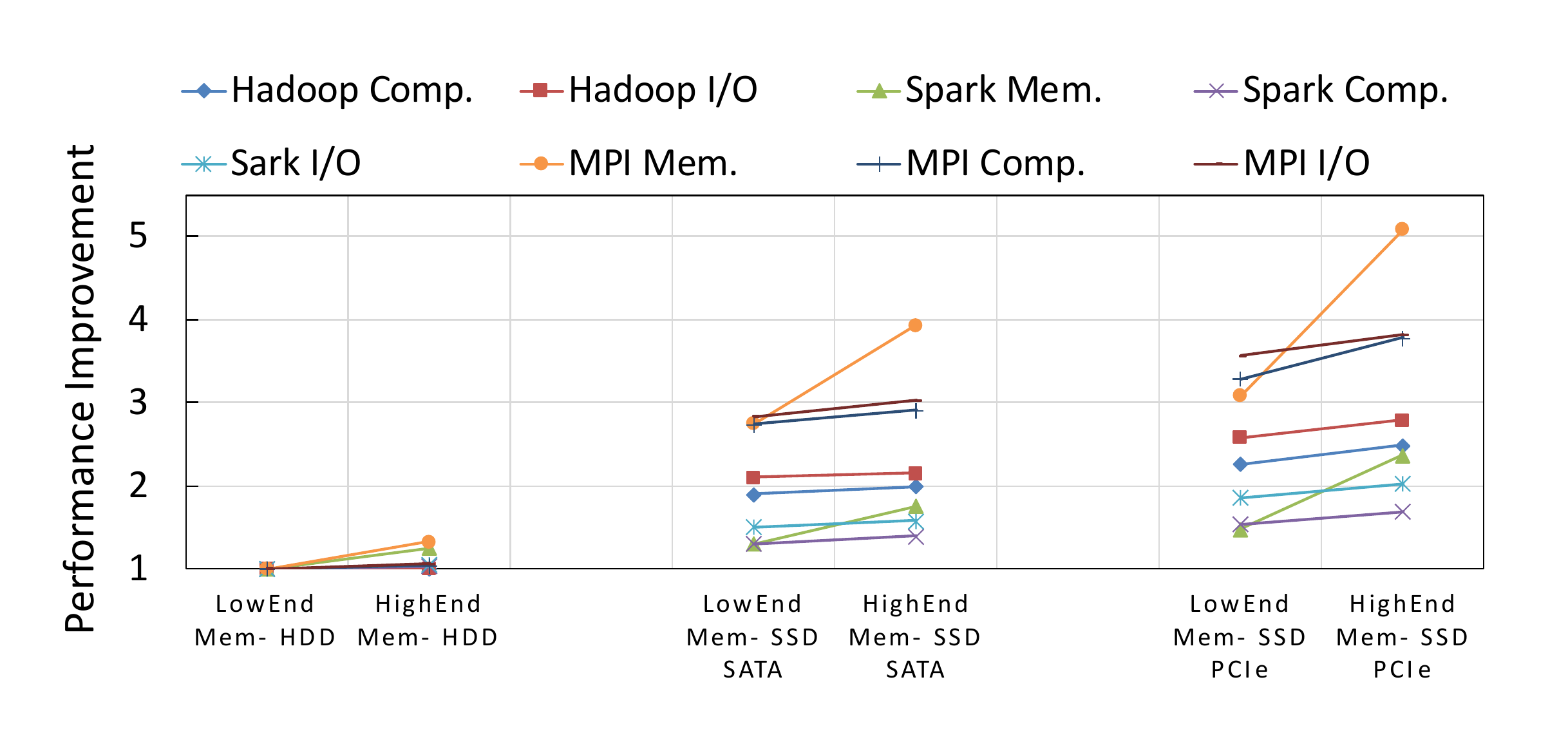}\\
\caption{Effect of memory and storage configuration on the performance}
\label{fig:disk}
\end{figure}

1) Disk implication: To show how choice of storage can change the performance while using different memory configuration, we performed several experiments using three types of storage (HDD, SSD SATA, and SSD PCIe). Figure \ref{fig:disk} shows that changing the disk from HDD to SSD PCIe improves the performance of Spark, Hadoop, and MPI by 1.6x, 2.4x, and 3.3x respectively. The reason that MPI workloads take more advantage from faster disk is that these workloads are written in C++. However, Hadoop and Spark are Java based frameworks and they use HDFS as an intermediate layer to access and manage storage. Our results show a high bandwidth DRAM is not required to accelerate the performance of MapReduce frameworks in presence of a slow HDD. However, MPI based workloads has the potential to benefit from high-end DRAM.

Another point regarding the storage is to use multiple disks per node to alleviate IO bottleneck. We performed a new set of experiments with two SSD storages per node. We found that HDFS is not aware of multiple disks on the node and all data is written or read from one disk. Therefore, using multiple disks per node does not guarantee the parallel access to the data blocks of HDFS to reduce the IO bottleneck. Another point is to use RAID. Since HDFS is taking care of fault-tolerance and "striped" reading, there is no need to use RAID underneath an HDFS. Using RAID will only be more expensive, offer less storage, and also be slower (depending on the concrete RAID configuration). Since the Namenode is a single-point-of-failure in HDFS, it requires a more reliable hardware setup. Therefore, the use of RAID is recommended only on Namenodes.
In our experiments, we also have increased aggregate storage by increasing the number of nodes (from 6 nodes to 12 nodes). We present the results in section III.C.3. The result reveals that increasing the number of nodes not only reduces IO requests’ pressure on each node, but also it reduces pressure on memory subsystem of each node.

It is important to note that SSD increases the read and write bandwidth of disk and substantially reduces the latency of access to disk compared to HDD. However, accessing to I/O means loosing of millions of CPU cycles, which is large enough to vanish any noticeable advantage of using a high-performance DRAM. On the other hand, the only way to take advantage of a SSD is to read or write a big file (hundreds of megabyte) at once but our result shows that HDFS reads the data in much smaller blocks, regardless of HDFS block size.

\begin{figure}[t]
\centering
 \includegraphics[width=0.48\textwidth]{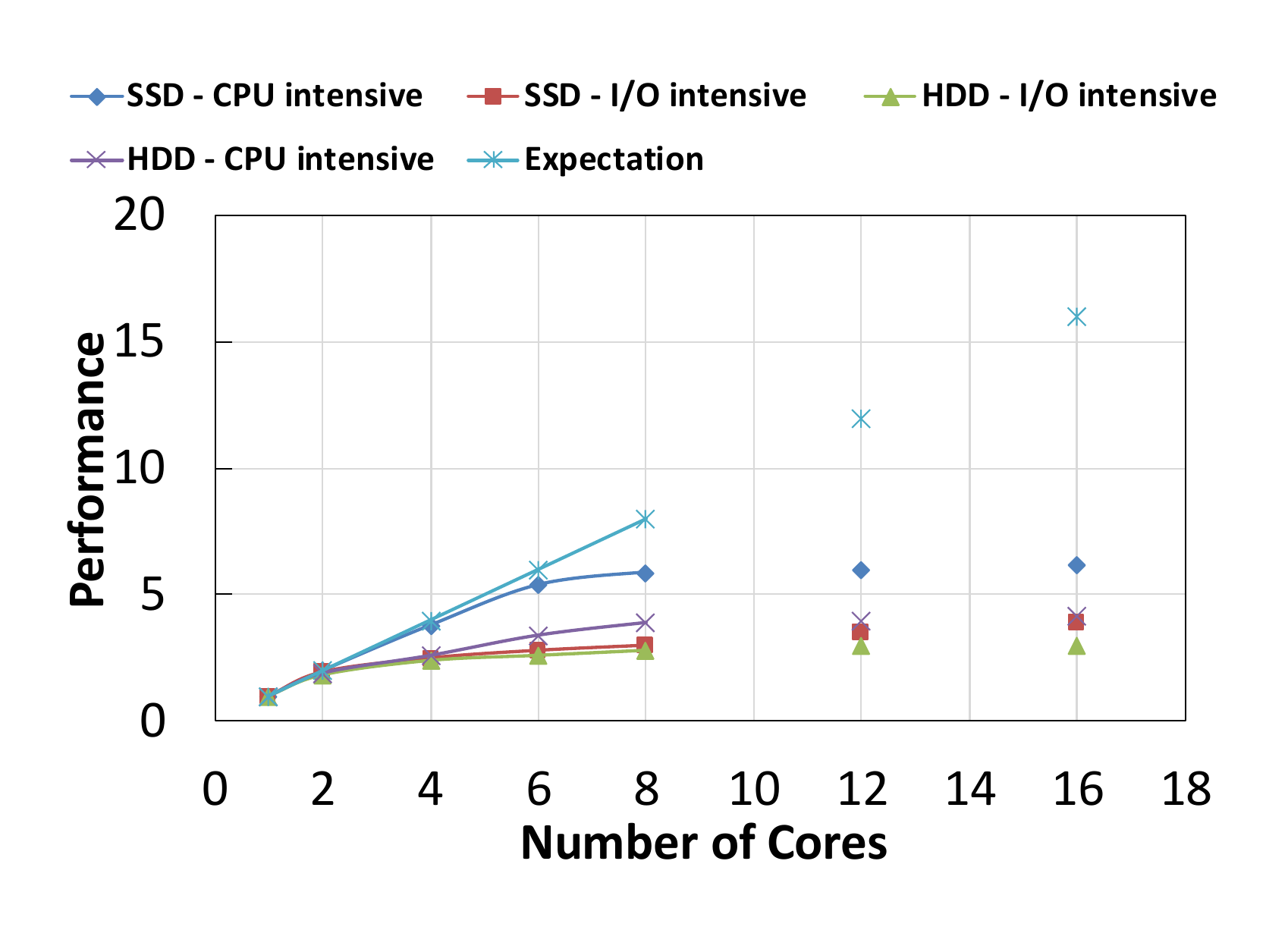}\\
\caption{Effect of core count on the performance}
\label{fig:core}
\end{figure}

2) Core count implication: In the previous section, we classified workloads into two groups of CPU intensive and I/O intensive. Figure \ref{fig:core} demonstrates the effect of increasing the number of cores per node on the performance of these two groups. The expectation is that performance of the system improves linearly by adding cores because big data workloads are heavily parallel. However, we observe a different trend. For CPU intensive workloads and when the core count is less than 6 cores per node, the performance improvement is close to the ideal case. The interesting trend is that increasing the number of cores per node does not improve the performance of data intensive workloads noticeably beyond 6 cores. As the increase in the number of cores increases the number of accesses to the disk, the disk becomes the bottleneck of the system. At 8 cores, the CPU utilization is dropped to 44\% for I/O intensive workloads, on average.

Based on these observations, we develop Eq. (\ref{eq:mc}) to find the number of cores for which further increase does not noticeably enhance the performance of system: 

\begin{eqnarray}\label{eq:mc}
\begin{aligned}
\centering
Max (cores) = ((BW \times Nd))/((Nsc \times Fr \times \lambda))   
\end{aligned}
\end{eqnarray}

We define the variables used in this equation as follow: BW is the nominal bandwidth of each disk. Nd is the number of disk installed on the server. Nsc is the number of sockets. Fr is CPU core frequency and Lambda is a constant, which we found through our real-system experiments. As the effective I/O bandwidth depends on the block size and I/O request size, we have used fio \cite{salah2011performance} to calculate Lambda for different block size requests, presented in figure 10. Designers can use this equation to select an optimum configuration, such as the number of cores, core frequency, disk type, and number of disks per node. As an example, the number of cores beyond which there is no noticeable performance gain on a server with one socket, one SSD storage with nominal 400 MBpS bandwidth, 16KB IO request size, running at 2.8 GHz is 8 based on the above equation. We have validated equation 1 for different classes of workload and the result is presented in Figure \ref{fig:table}. 

\begin{figure}[t]
\centering
 \includegraphics[width=0.48\textwidth]{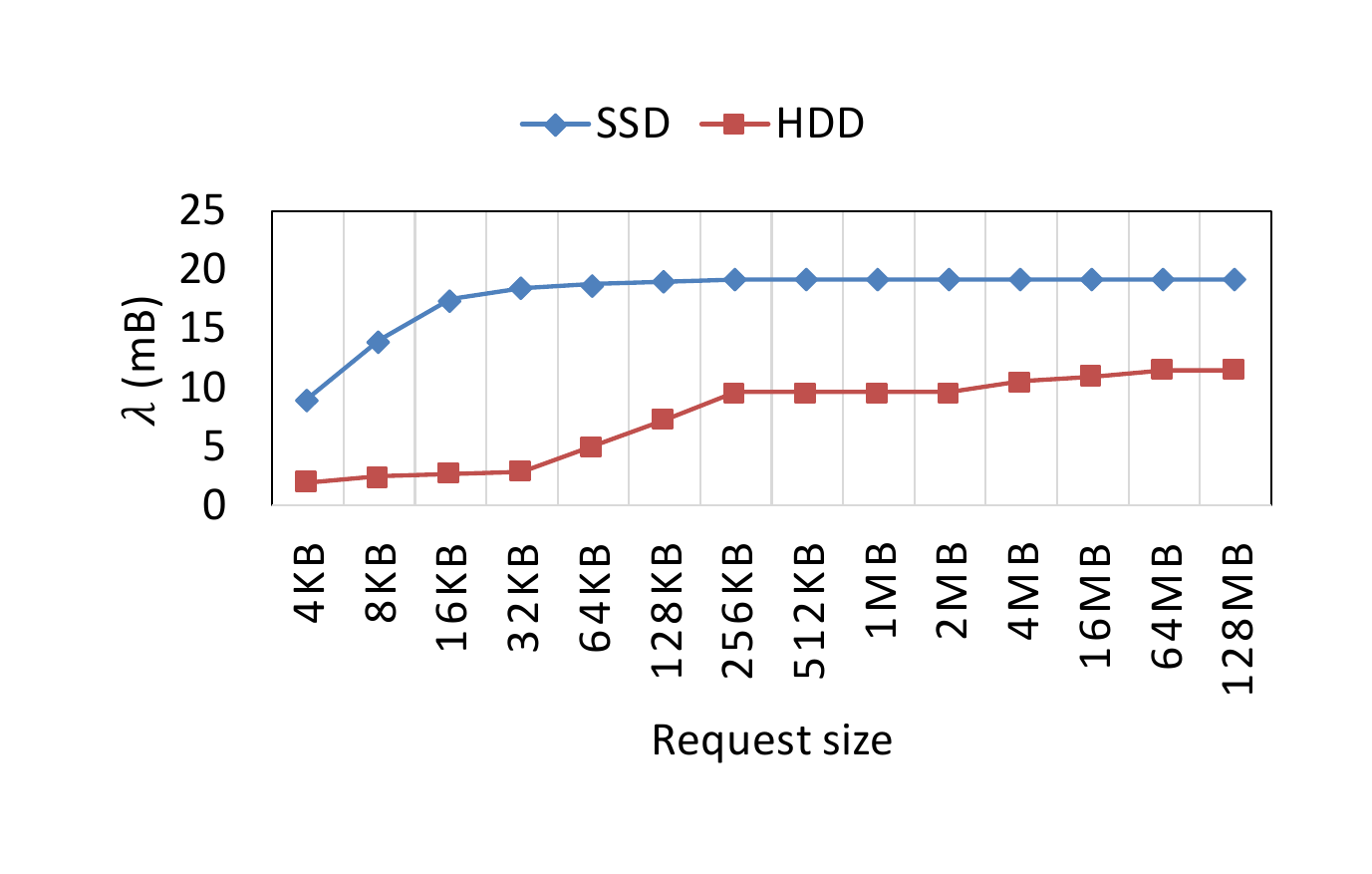}\\
\caption{Lambda value for different request size and storage type}
\label{fig:landa}
\end{figure}

\begin{figure}[t]
\centering
 \includegraphics[width=0.48\textwidth]{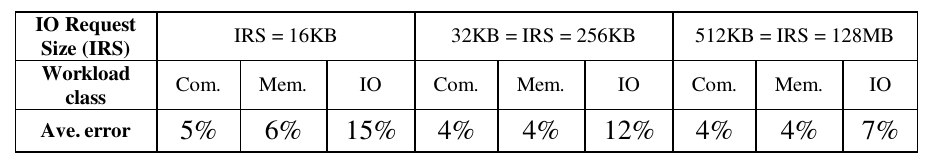}\\
\caption{Average error of optimum core count prediction}
\label{fig:table}
\end{figure}

\begin{figure}[t]
\centering
 \includegraphics[width=0.48\textwidth]{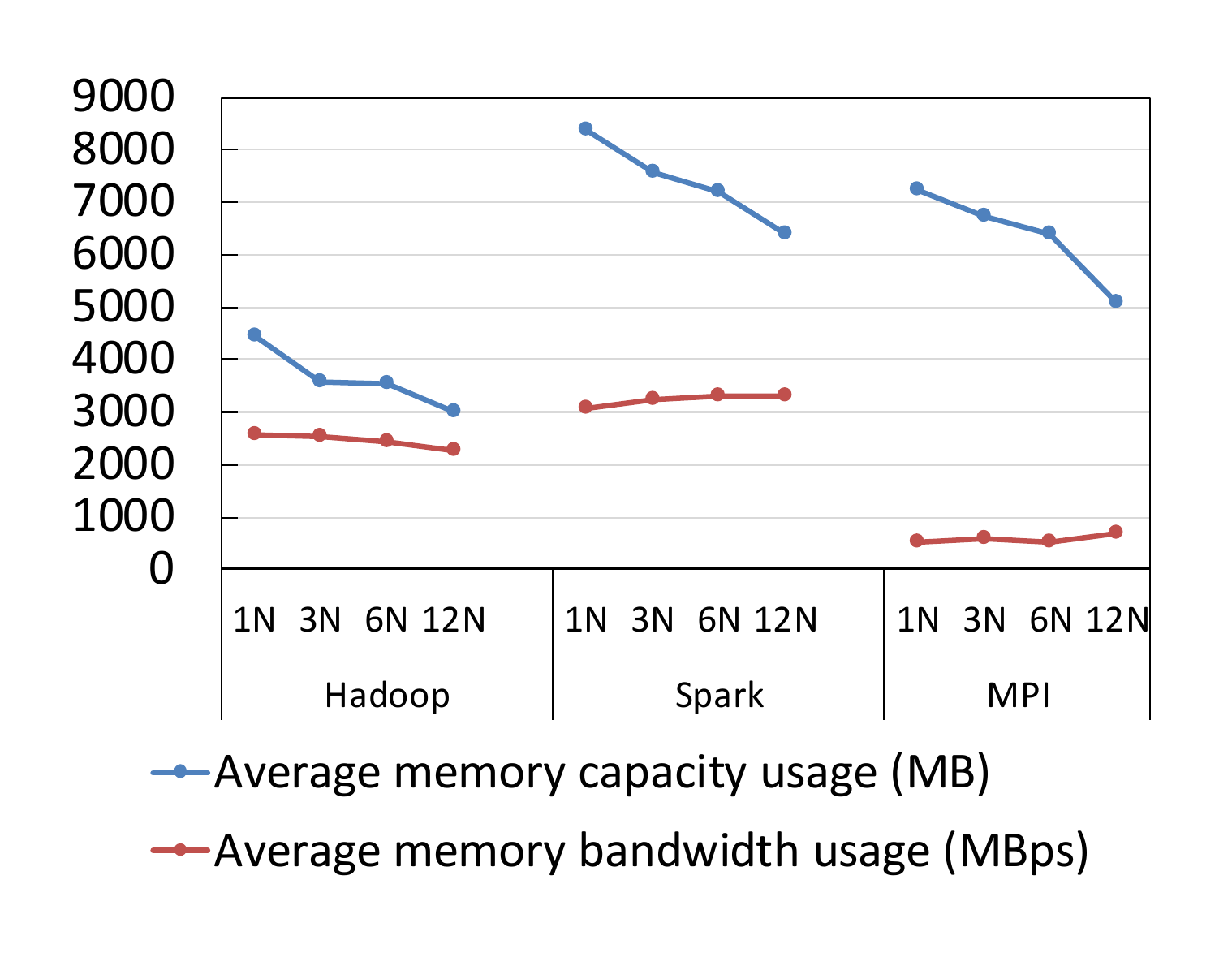}\\
\caption{Node count implication on memory usage}
\label{fig:cluster}
\end{figure}

3) Cluster size implication: Our cluster size (6 nodes) is small compared to a real-world server cluster. It is therefore important to understand the impact of cluster size on the memory characterization results. To achieve this, we performed three additional experiments with a single node, a three-node, as well as a twelve-node cluster to study the effect of cluster size on memory subsystem performance. Figure \ref{fig:cluster} shows the result of our experiments. These results show that increasing the size of cluster from 1 to 12 changes the memory behavior; it slightly reduces the pressure on memory subsystem. Increasing the cluster size reduces both memory usage and memory bandwidth utilization on each node, on average. Based on these results we anticipate that the memory subsystem mostly will not be under pressure in large scale server cluster. Therefore, we anticipate no urgent need to over provision memory subsystem in such environment for the studied workloads.

\section{Discussion}
\label{sec:dis}

The contribution of this paper is to give an insight on the role the memory and storage subsystem in the overall performance of the bare metal servers when running data intensive workloads. Our experimental results illustrate that data intensive workloads show three distinct behaviors of being I/O bound, compute bound, and memory bound. Based on the results presented in this paper, we observed that Hadoop framework is not memory intensive. This means Hadoop does not require high frequency, and large number of channels memory for higher performance. Moreover, Spark and Hadoop frameworks’ memory usage is bounded because they are MapReduce based frameworks and use HDFS block-based file system for storage. On the other hand, MPI framework’s memory usage is not predictable and different workload may have different demands for memory capacity. Our results show MPI and Spark based iterative tasks benefit from high frequency memory as they are memory bound workloads. Increasing the number of memory channels beyond two channels does not enhance the performance of those workloads. This is an indication for lack of efficient memory allocation and management in both hardware (memory controller) and software stacks. It is important to use our finding to help decision making in budgeting server infrastructure when building cloud or even when scheduling workloads. For instance, with a given budget to build a cluster, our results suggest allocating more budget to acquire more nodes rather than expensive high performance DRAM subsystems. Just as an example, the price of a server with two Sockets and 8 cores per socket with 128 GB memory running at 1333 MHz is equal to a server with one Socket and 8 cores with 256 GB memory running at 1866MHz.  For a designer who wants to select a server for MapReduce clusters, our results suggest the former one to be a better option.

\section{Related Work}
\label{sec:relat}

Latest works on memory characterization of big data applications are \cite{iiswc-makrani,igsc-makrani,ccgrid-makrani}. However, these works did not study the impact of storage. Another recent work studied the effect of memory bandwidth on the performance of MapReduce frameworks and presented a memory navigator for modern hardware \cite{MeNa}.
A recent work on big data \cite{dimitrov2013memory} profiles the memory access patterns of Hadoop and noSQL workloads by collecting memory DIMM traces using special hardware. This study does not examine the effects of memory frequency and number of channels on the performance of the system. A more recent work \cite{clapp2015quantifying} provides a performance model that considers the impact of memory bandwidth and latency for big data, high performance, and enterprise workloads. The work in \cite{alzuru2015hadoop} shows how Hadoop workload demands different hardware resources. This work also studies the memory capacity as a parameter that impacts the performance. However, as we showed in this work, their finding is in contrast with ours. In \cite{zhu2005performance} the authors evaluate contemporary multi-channel DDR SDRAM and Rambus DRAM systems in SMT architectures. The work in \cite{basu2013efficient} mainly focuses on page table and virtual memory optimization of big data and \cite{jia2013characterizing} presents the characterization of cache hierarchy for a Hadoop cluster. Few works \cite{depend,dft,ft-mem} studied the impact of fault tolerant techniques on the performance and memory usage of embedded system. These works do not analyze the memory subsystem.

A recent work on big data benchmarking \cite{xiong2013characterization} analyzes the redundancy among different big data benchmarks such as ICTBench, HiBench and traditional CPU workloads and introduces a new big data benchmark suite for spatio-temporal data. The work in \cite{pan2014characterization} selects four big data workloads from the BigDataBench \cite{wang2014bigdatabench} to study I/O characteristics, such as disk read/write bandwidth, I/O devices utilization, average waiting time of I/O requests, and average size of I/O requests. Another work \cite{liang2014performance} studies the performance characterization of Hadoop and DataMPI, using Amdahl’s second law. This study shows that a DataMPI is more balanced than a Hadoop system. In a more recent work \cite{hurt2015analysis} the authors analyze three SPEC CPU2006 benchmarks (libquantum, h264ref, and hmmer) to determine their potential as big data computation workloads. The work in \cite{beamer2015locality} examines the performance characteristics of three high performance graph analytics. One of their findings is that graph workloads fail to fully utilize the platform’s memory bandwidth.  The work in \cite{issa2016performance} performs performance analysis and characterizations for Hadoop K-means iterations. Moreover, there are studies on hardware acceleration of Hadoop applications that do not analyze the impact of memory and storage on the performance \cite{fccm,ss}. 

\section{Conclusions}
\label{sec:con}

Characterizing the behavior of data intensive workloads is important as it helps guiding scheduling decision in cloud scale architectures as well as helping making decisions in designing server cluster for big data computing. This work performs a comprehensive analysis of memory requirements through an experimental evaluation setup. We study diverse domains of workloads from microkernels, graph analytics, machine learning, E-commerce, social networks, search engines, and multimedia in Hadoop, Spark, and MPI. This gives us several insights into understanding the memory and storage role for these important frameworks. We observe that most of studied workloads in MapReduce based frameworks such as Hadoop and Spark do not require a high-end memory. On the other hand MPI workloads, as well as iterative tasks in Spark (e.g. machine learning) benefit from a high-end memory. Moreover, our result shows that changing the disk from HDD to SSD improves the performance of Spark, Hadoop, and MPI by 1.6x, 2.4x, and 3.3x respectively. However, I/O bandwidth caps the performance benefit of multicore CPU. Therefore, we developed an experimental equation to help designers to find the number of cores for which further increase does not enhance system performance noticeably. Moreover, we found that the current storage systems are the main bottleneck for the studied applications hence any further improvement of memory and CPU architecture without addressing the storage problem is a waste of money and energy. To the best of our knowledge this is the first work that looks beyond just the memory capacity to understand memory behavior by analyzing the effect of memory frequency as well as the number of memory channels on the performance of Hadoop, Spark, and MPI based big data workloads.

\begin{tiny}
\bibliographystyle{IEEEtran}
\bibliography{literature}
\end{tiny}

\end{document}